\documentclass[entropy,article,submit,moreauthors,pdftex,10pt,a4paper]{Definitions/mdpi} 

\usepackage{subfigure}
\usepackage{verbatim}
\usepackage{dcolumn}
\usepackage{bm}
\usepackage{epsf}
\usepackage{color}
\usepackage{dsfont}
\usepackage{amsmath,amssymb}
\usepackage{url}
\newcommand{\bra}[1]{\left\langle #1\right|}
\newcommand{\ket}[1]{\left|#1\right\rangle}
\newcommand{\braket}[2]{\left\langle #1|#2\right\rangle}

\newcommand{\la}{\left\langle}
\newcommand{\ra}{\right\rangle}
\newcommand{\pd}{\partial}

\newcommand{\id}{\mathbb{I}}

\newcommand{\bla}{bla\\bla\\bla\\bla\\bla}
\newcommand{\mb}[1]{\mbox{\boldmath$#1$}}
\newcommand{\mc}[1]{\mathcal{#1}}
\newcommand{\mbb}[1]{\mathbb{#1}}
\newcommand{\mf}[1]{\mathfrak{#1}}
\newcommand{\mrm}[1]{\mathrm{#1}}

\firstpage{1} 
\pubvolume{1}
\issuenum{1}
\articlenumber{0}
\pubyear{2021}
\copyrightyear{2021}
\history{Received: date; Accepted: date; Published: date}
\preto{\abstractkeywords}{\nolinenumbers}
\usepackage{titlesec}
\usepackage{physics}

\Title{Time-rescaling of Dirac dynamics: shortcuts to adiabaticity in ion traps and Weyl semimetals}

\Author{Agniva Roychowdhury $^{1}$* \orcidA{}, Sebastian Deffner $^{1,2}$ \orcidB{}}

\AuthorNames{Agniva Roychowdhury, Sebastian Deffner}

\address{$^{1}$ \quad Department of Physics, University of Maryland, Baltimore County, Baltimore, MD 21250, USA; deffner@umbc.edu \\
$^2$ \quad Instituto de F\'isica `Gleb Wataghin', Universidade Estadual de Campinas, 13083-859, Campinas, S\~{a}o Paulo, Brazil}

\corres{Correspondence: agniva.physics@umbc.edu}

\abstract{Only very recently, rescaling time has been recognized as a way to achieve adiabatic dynamics in fast processes. The advantage of time-rescaling over other shortcuts to adiabaticity is that  it does not depend on the eigenspectrum and eigenstates of the Hamiltonian. However, time-rescaling requires that the original dynamics are adiabatic, and in the rescaled time frame the Hamiltonian exhibits non-trivial time-dependence. In this work, we show how time-rescaling can be applied to Dirac dynamics, and we show that all time-dependence can be absorbed into the effective potentials through a judiciously chosen unitary transformation. This is demonstrated for two experimentally relevant scenarios, namely for ion traps and adiabatic creation of Weyl points.}


\begin{document}

\section{Introduction}

From the very beginning of quantum control, it has been recognized that circumventing the quantum adiabatic theorem \cite{Born1927} poses a formidable challenge. In essence, this theorem asserts that any quantum process, that is driven at rates larger than the typical energy gaps, is inevitably accompanied by excitations \cite{Messiah1966,Nenciu1980, Nenciu1980a, Nenciu1981}. In the real world, these parasitic excitations are often not only undesirable, but even detrimental. For instance, in adiabatic quantum computing \cite{Albash2018} finite-time effects constitute a major source for computational errors \cite{Young2013,Gardas2018}.  Thus, to circumvent, mitigate, and suppress such finite-time excitations in controlled quantum processes a wide variety of techniques has been developed. Among the most successful approaches are transitionless quantum driving \cite{Demirplak2003,Demirplak2005,Berry2009,delCampi2013,Deffner2014}, the fast-forward technique \cite{Masuda2010,Masuda2011,Masuda2014,Masuda2018}, and methods that rely on identifying the adiabatic invariants \cite{Chen2010,Torrontegui2014,Kiely2015,Jarzynski2017}, to name just a few. For a comprehensive exposition of the field ``shortcuts to adiabaticity'' we refer to recent reviews \cite{Torrontegui2013,Guery2019}, a special collection of articles \cite{delCampo2019}, and a perspective \cite{Deffner2020EPL}.

Somewhat naturally, the majority of work has focused on quantum processes that can be described by time-dependent Schr\"odinger equations. However, shortcuts to adiabaticty have also found generalizations and applications in, e.g., open system dynamics \cite{Alipour2020}, classical dynamics \cite{Patra2017,Patra2017NJP}, and even biologically relevant settings \cite{Iram2020}. Complementing these efforts, the present paper focuses on relativistic quantum dynamics. This is motivated by recent work that has generalized the fast-forward technique \cite{Deffner2015b}, transitionless quantum driving \cite{fan2017}, and invariant based methods \cite{song2017} to controlled Dirac dynamics.

The Dirac equation \cite{Dirac1928} was originally formulated to describe the properties of massive spin-$1/2$ particles, such as electrons and quarks \cite{Thaller1956,Peskin1995}. However, in recent years it attracted wider attention \cite{Pickl2008,Fillion2012,Fillion2013,Fillion2013a,Fillion2015,Villamizar2015,Schmidt2015,Deffner2015}, which is mostly motivated by the discovery of so-called Dirac materials \cite{Wehling2014}. In these systems the dispersion relation becomes linear, and hence the low-energy excitations behave more akin to massless Dirac particles than fermionic Schr\"odinger particles.

In the context of shortcuts to adiabaticity, the technical challenges already present for Schr\"odinger dynamics become significantly more involved for Dirac dynamics \cite{Faisal2011}. One way or another, implementing most shortcuts requires knowledge of the energy spectrum, or the use of highly non-local control fields. Therefore, any technique that requires less detailed information about the dynamics appears highly desirable \cite{Deffner2020EPL}. In the following, we propose and demonstrate how the method of ``time-rescaling'' \cite{Bernardo2020} is generalized to Dirac dynamics. We will see that while simply applying a scaling transformation is mathematically straight forward, the potential physical implementations are markedly less clear. This originates in the fact that rescaling time leads to an effectively  time-dependent mass \cite{Bernardo2020}. We will show how this can be remedied by a judiciously chosen unitary transformation of the Dirac equation. The experimental applicability of our findings are demonstrated for two relevant systems, namely for ion traps and adiabatic pumping in Weyl semimetals.

Due to the wide variety of concepts used in the following analysis, the narrative has been written as self-contained as possible. In Sec.~\ref{sec:pre} we summarize the main properties of the Dirac equation, and briefly review time-rescaling for Schr\"odinger dynamics. In Sec.~\ref{sec:time} we develop the method of time-rescaling for general Dirac dynamics. Sec.~\ref{sec:ion} is dedicated to adiabatically driving laser ion traps, and Sec.~\ref{sec:weyl} presents adiabatic pumping in Weyl semimetals. Finally, the analysis is concluded with a few remarks in Sec.~\ref{sec:con}.

\section{Preliminaries\label{sec:pre}}

We start by outlining notions and notations, and by briefly reviewing instrumental results from the literature.

\subsection{Relativistic quantum mechanics: the Dirac equation} 

The Dirac equation has its origin in an attempt to reconcile special relativity and quantum mechanics \cite{Dirac1928}. In its original inception and in first quantization it correctly describes the properties of massive spin-$1/2$ particles. It can be written in space representation as \cite{Thaller1956},
\begin{equation}
\label{eq01}
i\hbar\, \dot{\Psi}(\mb{x},t)=\left[\mb{\alpha}\cdot\left(-i\hbar c\, \nabla+ \mb{A}(\mb{x},t)\right)+\alpha_0\,mc^2+\id_4\,V(\mb{x},t)\right]\,\Psi(\mb{x},t)\,.
\end{equation}
Here, $\Psi(\mb{x},t)$ the 4-dimensional Dirac spinor, i.e., the wave function of a charged spin-$1/2$ particle with rest mass $m$ at position $\mb{x}=(x_1,x_2,x_3)$, and $c$ is the speed of light. As usual, we denote the derivative with respect to time by a dot. 

In covariant form the matrices $\mb{\alpha}=(\alpha_1,\alpha_2,\alpha_3)$ and $\alpha_0$ can be expressed as \cite{Peskin1995,Thaller1956},
\begin{equation}
\label{eq02}
\alpha^0=\gamma^0\quad\mathrm{and}\quad\gamma^0\, \alpha^k=\gamma^k\,.
\end{equation}
The $\gamma$-matrices are commonly written in terms of $2\times 2$ sub-matrices with the Pauli-matrices $\sigma_x, \sigma_y, \sigma_z$ and the identity $\id_2$ as,
\begin{eqnarray}
\label{eq03}
\gamma^0=\begin{pmatrix}\id_2&0\\0&-\id_2 \end{pmatrix}\quad&\gamma^1=\begin{pmatrix}0&\sigma_x\\-\sigma_x&0 \end{pmatrix}\quad
\gamma^2=\begin{pmatrix}0&\sigma_y\\-\sigma_y&0 \end{pmatrix}\quad&\gamma^3=\begin{pmatrix}0&\sigma_z\\-\sigma_z&0 \end{pmatrix}\,.
\end{eqnarray}
Finally, $\mb{A}(\mb{x},t)$ is the vector potential, and $V(\mb{x},t)$ is the scalar potential. The electric and magnetic fields, $\mb{E}(\mb{x},t) $ and $\mb{B}(\mb{x},t) $, are given by
\begin{equation}
\label{eq04}
\mb{E}(\mb{x},t)=-\nabla\, V(\mb{x},t)- \dot{\mb{A}}(\mb{x},t)\quad \mathrm{and}\quad \mb{B}(\mb{x},t)=\nabla \times \mb{A}(\mb{x},t)\,.
\end{equation}
Note that the Dirac equation is gauge invariant \cite{Peskin1995}, and we can thus choose mathematically convenient representations.

In the following it will also prove convenient to introduce the Dirac Hamiltonian $\mc{H}_D$, with which Eq.~\eqref{eq01} can be expressed in basis-independent form,
\begin{equation}
\label{eq:dirac}
i\hbar\, \ket{\dot{\Psi}(t)}=\mc{H}_D(t)\,\ket{\Psi(t)}\,.
\end{equation}
Hence, the Dirac equation \eqref{eq:dirac} becomes formally identical to the time-dependent Schr\"odinger equation. It is worth emphasizing, however, that the standard Schr\"odinger Hamiltonian is quadratic in momentum, whereas the Dirac Hamiltonian is linear. This is a direct consequence of the relativistic energy-momentum relation, and it will become instrumental in the following analysis. Moreover, for massive particles the Dirac Hamiltonian contains the rest energy, which is not present in pure spins systems or in Schr\"odinger dynamics. We will see in the following that when rescaling time this additional term requires special attention.

\subsection{Time-rescaling of Schr\"odinger dynamcis}

Time-rescaling was put forward by Bernardo in Ref.~\cite{Bernardo2020} as an alternative method for finding shortcuts to adiabaticity, that does not depend on the instantaneous eigenstates of the dynamics. To this end, Ref.~\cite{Bernardo2020} considers the time-dependent Schr\"odinger equation
\begin{equation}
\label{eq:schrodinger}
i\hbar \,\ket{\dot{\psi}(t)}=\left(\frac{p^2}{2 m}+V(x,t)\right)\,\ket{\psi(t)}=H(t)\,\ket{\psi(t)}\,,
\end{equation}
where $H(t)$ is the standard Hamiltonian. The solution of Eq.~\eqref{eq:schrodinger} can be expressed in terms of the unitary evolution operator,
\begin{equation}
\label{eq:U}
U(\tau)=\mc{T}_>\,\exp{-\frac{i}{\hbar}\int^\tau_0 dt\,H(t)}
\end{equation}
where $\mc{T}_>$ denotes time-ordering.

Time-rescaling is then nothing else but a transformation of the time variable in the exponent of the unitary evolution. We have
\begin{equation}
\label{eq:Urescale}
U(\tau)=\mc{T}_>\,\exp{-\frac{i}{\hbar}\int^{f^{-1}(\tau)}_{f^{-1}(0)} ds\, \dot{f}(s)\,H(f(s))}\,,
\end{equation}
where $f(t)$ is an arbitrary rescaling function. It is then easy to see that Eq.~\eqref{eq:Urescale} can be exploited as a shortcut to adiabaticity for  any $\left|f^{-1}(\tau)- f^{-1}(0)\right|\leq\tau$. If the original dynamics \eqref{eq:U} describes an adiabatic process, then Eq.~\eqref{eq:Urescale} achieves the same adiabatic dynamics in shorter time, for all rescaling functions $f(t)$ that obey the boundary conditions 
\begin{equation}
\label{eq:f_bound}
f^{-1}(0)=0, \quad f^{-1}(\tau)=\tau/a, \quad\text{and}\quad \dot{f}(0)=\dot{f}(\tau/a)=1
\end{equation}
where $a>1$ determines the ``time contraction factor'' \cite{Bernardo2020}.

The shortcoming of this technique is that in rescaled variables, the new Hamiltonian becomes $\widetilde{H}(t)\equiv \dot{f}(t)H(f(t))$. Generically,  this leads to a time-dependent mass,  which is not easy to realize in experimental scenarios.  In Appendix \ref{sec:app} we show how this can be remedied with the help of a canonical transformation.  In particular,  we show that time-rescaling and counterdiabtic driving are equivalent for scale-invariant problems \cite{Deffner2014}. 

In the following, we will generalize and analyze Eq.~\eqref{eq:Urescale} for the time-dependent Dirac equation \eqref{eq:dirac}.  Special focus will be put on experimental accessibility of the arising time-dependent terms.

\section{Time-rescaling of Dirac dynamics \label{sec:time}}

It is easy to see that the solution of the time-dependent Dirac equation \eqref{eq:dirac} can be expressed as
\begin{equation}
\mc{U}_D=\mc{T}_>\,\exp{-\frac{i}{\hbar}\int^\tau_0 dt\,\mc{H}_D(t)}\,.
\end{equation}
Hence the time-rescaled dynamics become
\begin{equation}
\mc{U}_D(\tau)=\mc{T}_>\,\exp{-\frac{i}{\hbar}\int^{f^{-1}(\tau)}_{f^{-1}(0)} ds\,\widetilde{\mc{H}}_D(s)} \,,
\end{equation}
where $\widetilde{\mc{H}}_D(t)=\dot{f}(t)\,\mc{H}_D(f(t))$ is the time-rescaled Dirac Hamiltonian.

Hence, it appears to be rather straight forward to employ time-rescaling as a shortcut to adiabaticity also in Dirac dynamics. That the situation is not quite that simple becomes apparent when inspecting the explicit form of the Dirac Hamiltonian \eqref{eq01}. Notice, that while time-rescaling Schr\"odinger dynamics only led to a time-dependent mass, for Dirac dynamics the time-scaled Hamiltonian $\widetilde{\mc{H}}_D(t)$ is governed by an effectively time-dependent speed of light, $\widetilde{c}(t)=\dot{f}(t)\, c$, and an effectively time-dependent \emph{rest} energy $\widetilde{m}(t) \widetilde{c}(t)^2=\dot{f}(t)\,m c^2$. 

We will see in the following Sec.~\ref{sec:ion} that considering $\widetilde{c}(t)$ and $\widetilde{m}(t)$ is perfectly reasonable and realizable in ion traps, which are described by an \emph{effective} Dirac equation. In general, however, it seems rather implausible that the speed of light and the rest energy can be considered time-dependent control parameters. Thus, we continue the analysis by proposing a canonical transformation that maps the effective time dependence exclusively onto vector and scalar potential, $\mb{A}(\mb{x},t)$ and $V(\mb{x},t)$.

For the sake of simplicity and without loss of generality, we continue by considering a system that is restricted to the $x$-direction. In this case the 4-component Dirac spinor can be separated into two identical 2-component bispinors. For mathematical convenience, we choose a representation in which the $(1+1)$-dimensional Dirac-equation reads \cite{deCastro2003,Solomon2010},
\begin{equation}
\label{eq:H_Dirac}
\widetilde{\mc{H}}_D(t)=\left[\widetilde{c}(t)\,p+\widetilde{A}(x,t)\right]\sigma_x+\widetilde{m}(t) \widetilde{c}(t)^2\,\sigma_z+\widetilde{V}(x,t) \,\mbb{I}_2\,,
\end{equation}
and where we introduced the time-rescaled potentials $\widetilde{A}(x,t)=\dot{f}(t) A(x,t) $ and $\widetilde{V}(x,t)= \dot{f}(t) V(x,t)$.

\subsection*{Absorbing the time-dependence into the potentials}

Our goal is now to find a unitary transformation that allows to write the time-rescaled Hamiltonian in standard form, i.e., with time-\emph{in}dependent speed of light and rest mass. Arguably the simplest ansatz is given by
\begin{equation}
\label{eq:K}
K(t)=\exp{i\phi(t)\,\sigma_x}=\cos(\phi(t))\,\mbb{I}_2+i\sin(\phi(t))\,\sigma_x\,.
\end{equation}
Thus, we obtain
\begin{equation}
\begin{split}
&\mf{H}_D(t)=K^{\dagger}(t)\,\widetilde{\mc{H}}_D(x,p,t)\, K(t)-i\hbar\, K^{\dagger}(t)\,\pd_t\,K(t)\\
&\quad=\left[\widetilde{c}(t)\,p+\widetilde{A}(x,t)+\dot{\phi}(t)\right]\,\sigma_x+\widetilde{m}(t)\widetilde{c}(t)^2\,\cos(2\phi(t))\,\sigma_z+\widetilde{m}(t)\widetilde{c}(t)^2\,\sin(2\phi(t))\,\sigma_y+\widetilde{V}(x,t)\,\mbb{I}_2\,,
\end{split}
\end{equation}
and the corresponding solution is given by $\Phi(x,t)=K(t)\,\Psi(x,t)$. We immediately observe that the time-dependent rest energy is multiplied by $\cos(2\phi(t))$, whereas the kinetic term is modified by $\dot{\phi}(t)$. Therefore, the two effectively time-dependent quantities, $m(t)$ and $c(t)$, can be fully described by $\phi(t)$ and its derivative $\dot{\phi}(t)$, respectively.

We start by choosing
\begin{equation}
\cos(2\phi(t))=1/\dot{f}(t)
\end{equation}
for which the rest energy becomes time-independent. In complete analogy to the Schr\"odinger case \cite{Bernardo2020}, the time-dependence of the kinetic term can then be absorbed into the vector potential. In particular, we define
\begin{equation}
\mf{A}(x,p,t)\equiv \widetilde{A}(x,t)+(\dot{f}(t)-1)\,c\,p+\frac{\ddot{f}(t)}{2 \dot{f}(t)\,\sqrt{\left[\dot{f}(t)\right]^2-1}}\,,
\end{equation}
which is nothing else but $\widetilde{A}(x,t)$ expressed in the corresponding interaction picture plus a position independent term.

Thus, we are only left with the pseudoscalar term \cite{Haouat2007,Haouat2008}, that is proportional to $\sigma_y$. Pseudoscalar potentials correspond physically to driving the system with circularly polarized light. Now, defining a new (scalar) potential
\begin{equation}
\mf{V}(x,t)\equiv mc^2\,\sqrt{\left[\dot{f}(t)\right]^2-1}\,\sigma_y+\widetilde{V}(x,t)\,\mbb{I}_2\,,
\end{equation}
we finally obtain
\begin{equation}
\mf{H}_D(t)=\left[c p+\mf{A}(x,p,t)\right]\,\sigma_x+m c^2\,\sigma_z+\mf{V}(x,t)\,,
\end{equation}
where all the time-dependence has been absorbed into the potentials. At first glance, the momentum dependent vector potential may look unphysical. However, this is simply a consequence of the fact that the system is non-conservative, and hence the driven system will experience an intertial force \cite{Gardas2016PRA} due to the ``acceleration'' from the time-rescaling.

In conclusion, we have shown that while in Dirac dynamics we have two, instead of only one, effectively time-dependent parameters, a simple unitary transformation allows to absorb the time-dependence entirely into vector and scalar potentials. The resulting vector potential is fully analogous to what was proposed in Ref.~\cite{Bernardo2020} for Schr\"odinger dynamics, and the scalar potential contains a simple pseudoscalar term.

\section{Dirac dynamics in laser ion-traps \label{sec:ion}}

It has been experimentally demonstrated that under certain conditions ions in laser traps can be described by effective Dirac dynamics \cite{Leibfried2003,Lamata2007,Gerritsma2010,Muga2016NJP}. In general, the applied laser field couples  internal vibrational levels of the ion and its motional degrees of freedom. Hence, the total Hamiltonian reads,
\begin{equation}
H_\mrm{tot}=H_\mrm{m}+H_\mrm{e}+H_\mrm{int}
\end{equation}
where $H_\mrm{m}$ and $H_\mrm{e}$ describe motional and electronic degrees of freedom, respectively.

The interaction Hamiltonian $H_\mrm{int}$ can be written in (1+1) dimensions as \cite{Lamata2007}
\begin{equation}
H_\mrm{int}(t)=2\eta\, \Delta\,\gamma\, [p-A(t)] \,\sigma_x+\hbar\omega\, \sigma_z
\label{iontrap1}
\end{equation}
where  $\eta=k\sqrt{\hbar/2m\nu}$ is the Lamb-Dicke parameter, $m$ denotes the mass of the trapped ion, and $\nu$ is the axial frequency of the confining Paul trap. Note that the effective interaction can be tuned by applying an external magnetic field described by $A(t)$.  Further, $\Delta=\sqrt{\hbar/2m\nu}$ and is the width the ground-state wave-function, and $\gamma$ is the strength of the interaction, which is varied by modulating the laser source. Finally, $\omega$ describes the detuning of the laser and the resonance frequency of the two-level atom. In principle, both $\gamma$ as well as $\omega$ can be controlled externally, and varied as a function of time.

We immediately recognize $H_\mrm{int}$ as the Dirac Hamiltonian \eqref{eq:H_Dirac} for vanishing vector and scalar potentials, and  \cite{Lamata2007} 
\begin{equation}
\widetilde{c}(t)=2\eta\, \Delta\,\gamma(t)\quad\text{and}\quad \widetilde{m}(t)\widetilde{c}(t)^2=\hbar\omega(t)\,.
\end{equation}
Hence, $H_\mrm{int}$ is already in the time-dependent form required to implement a shortcut to adiabaticity by means of time-rescaling.

\subsection*{A simple demonstration}

Before we continue, it is instructive to demonstrate time-rescaled Dirac dynamics in ion traps with a simple example. To this end, consider a simple scenario, in which we again choose $V(x,t)=0$.  In interaction picture,  we have
\begin{equation}
\label{eq:dirac_simple}
\mc{H}_D(p,t)=H_\mrm{int}(t)=\left[p-\sin^2(\pi t/2\tau)\right]\,\sigma_x+\cos^2(\pi t/2\tau)\,\sigma_z
\end{equation}
Note, that the latter Hamiltonian is written in instantaneous units of $2\eta\, \Delta\,\gamma(t)$. Thus, we simply have $A(t)=\sin^2(\pi t/2\tau)$ and $\hbar\omega(t)=\cos^2(\pi t/2\tau)$. Its instantaneous eigenstates can be expressed as
\begin{equation}
\label{eq:PHI}
\ket{\Phi(t)}=\cos(\theta_t/2)\ket{0}+\sin(\theta_t/2)\ket{1}
\end{equation}
where $\ket{0}$ and $\ket{1}$ are basis states of $\sigma_z$ and $\theta_t=\text{Arctan}[p_t-A(t)/\hbar\omega(t)]$, where $p_t$ is the instantaneous momentum. We immediately observe that for large enough $\tau$ this simple Dirac Hamiltonian \eqref{eq:dirac_simple} describes adiabatic dynamics, i.e., a system prepared in an initial eigenstate will remain in the corresponding, instantaneous eigenstate.

Now let us assume that the desired target state of the dynamics is $\ket{\Phi_\mrm{target}}=(1/\sqrt{2})(\ket{0}+\ket{1})$, which is the instantaneous eigenstate of $\mc{H}_D(p,t)$ at $t=\tau$. Using time-rescaling this target state can be reached in shorter time. To this end, we choose the rescaling function $f(t)$ \cite{Bernardo2020} to read
\begin{equation}
\label{eq:f}
f(t)=a\,t-\frac{\tau\,(a-1)}{2\pi a}\,\sin\left(\frac{2\pi at}{\tau}\right)
\end{equation}
where as before $a\geq 1$ is the acceleration factor. It is easy to see \cite{Bernardo2020} that this choice fulfills all boundary conditions \eqref{eq:f_bound}, and notice that for $a=1$ we simply have $f(t)=t$, i.e., we recover the original dynamics.

\begin{figure}
\centering
\includegraphics[width=.8\textwidth]{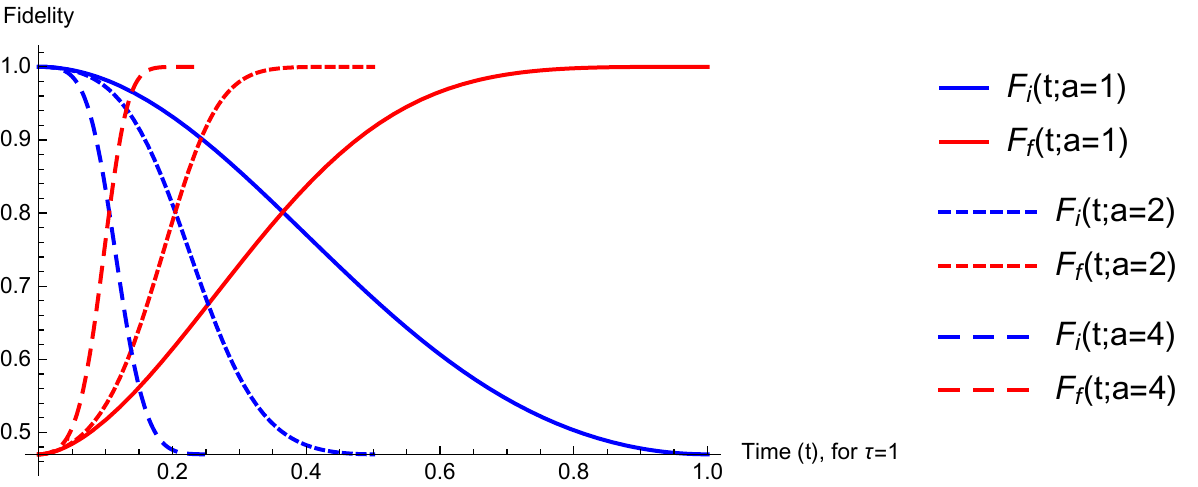}
\caption{\label{fig1} Fidelity of the time-evolved state with respect to initial and target state \eqref{eq:fid} for $a=1,2,4$ and $\tau=1$. Other parameters are set such that $\hbar =c=1$.}
\end{figure}

It is then a simple exercise to solve the dynamics numerically for the original Hamiltonian, $\mc{H}_D(t)$, \eqref{eq:dirac_simple} and its time-rescaled companion, $\widetilde{\mc{H}}_D(t)=\dot{f}(t)\, \mc{H}_D(t)$. To illustrate our findings, we choose the system to be initially prepared in $\ket{\Phi(0)}$ \eqref{eq:PHI},  and compute the fidelity of the time-evolved state with respect to initial and target state, $F_i(t)$ and $F_f(t)$, respectively. These are given by
\begin{equation}
\label{eq:fid}
F_i(t)=\left|\int^{+\infty}_{-\infty}dp\, \braket{\Psi(p,t)}{\Phi(0)}\right|^2\quad\text{and}\quad F_f(t)=\left|\int^{+\infty}_{-\infty}dp\,\braket{\Psi(p,t)}{\Phi_\mrm{target}}\right|^2\,.
\end{equation}
In Fig.~\ref{fig1} we depict the fidelities for a range of values of $a$. As expected, we see that for $a>1$ the target state is reached in time $\tau/a$.

\section{Time-rescaling Weyl semimetals: shortcuts to adiabatic pumping \label{sec:weyl}}

As a second example we discuss time-rescaling in the context of adiabatic pumping for Weyl semimetals. To this end, we need to apply the above developed framework to Floquet theory, first.

\subsection{Floquet theory and time-rescaling}

The solution of periodically driven many-body Hamiltonians can be expanded in the so-called Floquet states \cite{Floquet1883}. To this end, consider a general Hamiltonian \cite{Lindner2011},
\begin{equation}
H(t)=\sum_kH_{nm}(\mathbf{k},t)a_{n,\mathbf{k}}a^\dagger_{m,\mathbf{k}}+h.c.
\end{equation}
where $\mb{k}$ is the wave vector in the first Brilloin zone. Here, $a$ and $a^\dagger$ are the fermionic annihilation and creation operators, and $n$ and $m$ are indices associated with the system's degrees of freedom. If this Hamiltonian is periodic in time, $H(t)=H(t+T)$, the single particle wave function can be written as
\cite{Shirley1965,Floquet1883},
\begin{equation}
\label{eq:floq}
\psi(\mathbf{k},t)=\exp{-\frac{i}{\hbar}\,E(\mathbf{k})t}\,\phi(\mathbf{k},t)
\end{equation}
where $\phi(\mathbf{k},t)=\phi(\mathbf{k},t+T)$ is the Floquet mode with  periodicity $T$. The quasienergies $E(\mathbf{k})$ are  eigenvalues of the Floquet operator equation, and they satisfy,
\begin{equation}
(H(\mathbf{k},t)-i\hbar\,\partial_t)\,\phi(\mathbf{k},t)=E(\mathbf{k})\,\phi(\mathbf{k},t).
\end{equation}
Hence, the $E(\mathbf{k})$ exhibit the corresponding Floquet band structure, if the $E(\mathbf{k})$ are also periodic in momentum space. Interestingly, for fermionic systems Floquet theories predicts a variety of topological states of matter \cite{Lindner2011}, which we will be exploiting in the following.

The major advantage of the Floquet ansatz \eqref{eq:floq} is that the time-evolution operator can be factorized. We have
 \cite{Hanggi1998},
\begin{equation}
\label{eq:floq_op}
U_F(nT,0)=\mathcal{T}_>\,\prod^n_{k=1}\exp{-\frac{i}{\hbar}\int^T_0dt\,H(t)}
\end{equation}
and $\ket{\psi(nT)}=U_F(nT,0)\,\ket{\psi(0)}$. Thus, it is not difficult to realize that time-rescaling can be used to shorten the periodicity of the solutions, and in particular we can write
\begin{equation}
U_F(nT,0)=\mathcal{T}_>\prod^n_{k=1}\exp{-\frac{i}{\hbar}\int^{f^{-1}(T)}_{f^{-1}(0)} ds\, \dot{f}(s)H(f(s))}
\end{equation}
where, as above, $s$ is the re-scaled time variable and $\dot{f}(s)$ is the re-scaling function.

If we again consider time-acceleration, i.e., $\left|f^{-1}(T)-f^{-1}(0)\right|=T/a$ for $a>1$, then the periodicity of the Floquet states is shorted by a factor of $1/a$. In the following, we will exploit this observation for adiabatic evolution between Floquet points, or in other words will we use time-rescaling as a shortcut to adiabatic pumping.

In this context, it is also interesting to note that typically a periodically driven Hamiltonian does not have to be periodic in momentum space. However, it is still possible to introduce the Floquet operator \label{eq:floq_eq} in terms of a periodic momentum transport variable \cite{Ho2012}. Experimentally, this transport variable can be manifested by phase-shifting two optical-lattice potentials in a ratchet accelerator (RA) model \cite{Ho2012}, and its  infinitesimal evolution  corresponds to tracing out the entire Floquet band topology, by moving between different eigenstates of the Floquet operator.

\subsection{Shortcut to adiabatic creation of Weyl points}

Only rather recently, it has been shown that Floquet points can exhibit linear dispersion relations. For instance, Refs.~\cite{bm16a,bm16b} found topological phases that correspond to Weyl semimetals with three-dimensional Dirac ``cones''. This was achieved, by considering a modified, extended kicked Harper model and careful tuning of ``hopping'' and ``kicking'' strengths. 

We will now briefly outline how time-rescaling can be applied to create these Weyl points in Floquet theory in shorter time. To this end, we consider the off-diagonal kicked Harper model, whose Hamiltonian reads \cite{bm16b}
\begin{equation}
H(t)=\sum^{N-1}_{n=1}\left\{\left[J+(-1)^n\lambda\cos(\phi_y)\right]\ket{n+1}\bra{n}+h.c.\right\}+\sum^{N-1}_{n=1}\sum_{j}(-1)^n\left[V_1+V_2\cos(\Omega t)\right]\cos(\phi_z)\ket{n}\bra{n}
\end{equation}
where $n$ is the lattice site index, $J$ and $\lambda$ are parameters controlling the hopping strength, $N$ is the total number of lattice sites. Further, $V_1$ is the onsite potential, $V_2$ represents the coupling with the harmonic driving field, and $\Omega=2\pi/T$. As before, $T$ is the period of driving, and $H(t)=H(t+T)$. In this model \cite{bm16b}, $\phi_y$ and $\phi_z$ are quasimomenta, that can take any value in $(-\pi,\pi]$. Therefore, we can simplify the analysis again to (1+1) dimensions. It is also interesting to note, that Weyl chirality can be manifested by observing the time-evolution of the mean position, $\la\Delta x\ra$. This is facilitated by adiabatic transport in momentum space \cite{Ho2012}, or rather by adiabatic variation of $\phi_y$ and $\phi_z$. Thus, we focus now on applying time-rescaling for adiabatic transport to observe the Weyl chirality in $\la\Delta x\ra$.

For a single mode $k$, we have \cite{bm16b}
\begin{equation}
H_k(t)=2J\cos(k)\sigma_x+2\lambda \sin(k)\cos(\phi_y)\sigma_y+\left[V_1+V_2\cos(\Omega t)\right]\cos(\phi_z)\sigma_z\,.
\label{weyl1}
\end{equation}
It can be shown that the corresponding quasienergy spectrum exhibits band touching  points with linear dispersion relation \cite{bm16a,bm16b}.

For time-dependent variation of $\phi_y(t)$ and $\phi_z(t)$ Eq.~\eqref{weyl1} can be written as \cite{bm16b}
\begin{equation}
\begin{split}
&\hat{H}_k(t)=\left[2J\cos(k)\cos(2\alpha)+2\lambda\sin(k)\cos(\phi_y(t))\sin(2\alpha)\right]\,\sigma_x\\
&\quad+\left[-2J\cos(k)\cos(2\alpha)+2\lambda\sin(k)\cos(\phi_y(t))\sin(2\alpha)\right]\,\sigma_y+V_1\cos(\phi_z(t))\,\sigma_z
\end{split}
\end{equation}
where $\alpha=V_2\cos(\phi_z(t))\sin(\Omega t)/\hbar\Omega$. The corresponding time-rescaled Hamiltonian becomes, $\widetilde{H}_k(t)\equiv \dot{f}(t)\hat{H}_k(t)$, and we now need to verify that also $\widetilde{H}_k(t)$ exhibits the desired Weyl chirality.

In close proximity of the band-touching bounds \cite{bm16b} we can write
\begin{equation}
\widetilde{H}_k(t)\simeq \dot{f}(t)\left\{H_\mrm{pert}+\left[\ell\pi-V_1k_z\sin(\phi_l)\right]\,\sigma_z\right\}
\label{weyl2}
\end{equation}
where we introduced \cite{bm16b}
\begin{equation}
\begin{split}
H_\mrm{pert}&=\left\{-2Jk_x\,\cos[\ell c\,\sin(\Omega f(t))]-2\lambda k_y\,\sin[\ell c\,\sin(\Omega f(t))]\right\}\,\sigma_x\\
&\quad+\left\{2Jk_x\,\sin[\ell c\,\sin(\Omega f(t))]-2\lambda k_y\,\cos[\ell c\,\sin(\Omega f(t))]\right\}\,\sigma_y\,.
\end{split}
\end{equation}
Moreover, we have  $k_x=k-\pi/2$, $k_y=\phi_y-\pi/2$, $k_z=\phi_z-\phi_l$, $c=V_2/V_1$ and $\phi_l=\cos^{-1}(\ell \pi/V_1)$. Finally, $\ell$ denotes the quantum number of the quasienergy. Adiabatic momentum transport is then achieved by parametrizing $\phi_y$ and $\phi_z$ according to $\phi_y=\phi_{y,0}+r\cos(\theta(t))$ and $\phi_z=\phi_{z,0}+r\sin(\theta(t))$ and evolving $\theta(t)$ over a time period $T_0\gg T$ \cite{bm16b}. Thus, it is worth emphasizing that the relevant Floquet operator quantifies this periodicity, $T_0$, in momentum space, and not the time period of the driving field $T$.

Solving for the time-evolution operator exactly is hardly feasible. However, the Floquet operator \eqref{eq:floq_op} can be obtained from time-dependent perturbation theory. It can be shown \cite{bm16b} that we have for the first period
\begin{equation}
\label{eq:Ufloq}
U_F(T_0,0)\simeq \mbb{I}+i\left(2Jk_x\,\sigma_x+2\lambda k_y\,\sigma_y\right)\mc{J}(\ell c)
\end{equation}
where $\mc{J}$ is the Bessel function of the first kind. It is then a simple exercise to show that we obtain for the time-rescaled Floquet operator
\begin{equation}
\widetilde{U}_F(T_0/a,0)\simeq \mbb{I}+i\left(2Jk_x\,\sigma_x+2\lambda k_y\,\sigma_y\right)\mc{J}(\ell c)
\end{equation}
which is identical to the Floquet operator \eqref{eq:Ufloq}. However, due to time-rescaling $\widetilde{U}_F$ has a periodicity of $T_0/a$, which for $a>1$ described sped-up dynamics. Since the time-rescaled Floquet operator is identical to the original $U_F(T_0,0)$ all further steps of the analysis in Ref.~\cite{bm16b} remain true, yet the Weyl chirality is obtained with periodicity $T_0/a$.

Finally, we briefly remark on the complexity of the time-dependence in the time-rescaled dynamics. As before, in $\widetilde{H}_k(t)$ all original parameters are multiplied by $\dot{f}(t)$. In particular, this requires both kicking and hopping strengths to be varied with time. However, it is not hard to see that in complete analogy to above in Sec.~\ref{sec:time} the time-dependence can be absorbed into potentials. This can be facilitated again with, e.g., the simplest unitary transformation $K(t)$ \eqref{eq:K}.

\section{Concluding remarks \label{sec:con}}

Controlling quantum systems is a ubiquitous goal in the development of quantum technologies. To this end, shortcuts to adiabaticity provide a powerful tool kit to steer quantum system towards desired target states. However, most techniques are rather complicated to be implemented in realistic scenarios, since most of them require exquisite knowledge about the eigenspectrum of the driven Hamiltonians. 

Time-scaling relies on the  simple idea that rather than controlling single states, faster dynamics can be achieved by simply transforming the dynamics to a new time-frame. To utilize time-rescaling as a shortcut to adiabaticty, two criteria need to me met: (i) the original dynamics is adiabatic, and (ii) the resulting Hamiltonian has to be realizable and physical. 

Using time-rescaling as a shortcut was originally proposed only for Schr\"odinger dynamics. The natural question arose, whether also relativistic dynamics can be treated in this framework. To answer this question we have analyzed Dirac dynamics in first quantization. In second quantization, one would have to be concerned with pair production and radiation, and hence a quantum field theoretic description would become necessary. 

In the present paper we have shown that time-rescaling can be directly applied to Dirac dynamics, and that the aforementioned criteria are met by at least two experimentally relevant scenarios, namely laser ion traps and adiabatic creation of Weyl points. Thus, we remain optimistic that time-rescaling will, indeed, find applications in experimental settings

\vspace{6pt}

\authorcontributions{Both authors contributed equally to this study.}

\funding{This research was supported by grant number FQXi-RFP-1808 from the Foundational Questions Institute and Fetzer Franklin Fund, a donor advised fund of Silicon Valley Community Foundation (SD).}

\acknowledgments{ARC thanks Pierre Naz\'e for bringing Ref. \cite{Bernardo2020} to our attention. Enlightening discussions with Akram Touil and Nirnoy Basak are gratefully acknowledged.}

\conflictsofinterest{The authors declare no conflict of interest.}

\appendixtitles{yes} 
\appendix

\section{Shortcuts to adiabaticity from time-rescaling in scale invariant problems \label{sec:app}}

This appendix is dedicated to the question whether time-rescaling is related, or even equivalent to other techniques for shortcuts to adiabaticity. To this end, we analyze the method for scale-invariant problems, in both classical and quantum dynamics.

\subsection{Classical Hamiltonian dynamics}

We start by considering a classical system that evolves under Hamiltonian, scale-invariant dynamics. For such problems it has been shown that the counterdiabatic field facilitating transitionless quantum driving takes a particularly simple form \cite{Deffner2014}. 

The corresponding Hamiltonian is given by
\begin{equation}
H(x,p,t)=\frac{p^2}{2m}+\frac{1}{\gamma^2}V\left(\frac{x}{\gamma}\right)
\end{equation}
where $m$ is the mass of the particle, $\gamma$=$\gamma(t)$ and $V$ is the scale-invariant driving potential obeying $V(x,t)=V(x/\gamma)$. Note that time-rescaling can be applied directly to classical dynamics, by replacing the commutator in the Heisenberg equation of motion with the Poisson bracket.

The corresponding time-rescaled Hamiltonian becomes
\begin{equation}
\widetilde{H}(x,p,t)=\dot{f}(t)\,\frac{p^2}{2m}+\frac{\dot{f}(t)}{\gamma(f(t))^2}\,V\left(\frac{x}{\gamma(f(t))}\right)
\end{equation}
which has the same technical problem that we discussed in the main text, namely $\widetilde{H}(x,p,t)$ contains an effectively time-dependent mass. Scaling time is actually a well-studied problem in Hamiltonian dynamics, and it can be expressed in terms of an infinitesimal canonical transformation \cite{Carinena1988}.

Therefore, we now consider a generating function of the second kind \cite{Goldstein1980}
\begin{equation}
F(x,\bar{p},t)=h_1(t)\,x\bar{p}+h_2(t)\,x^2
\end{equation}
where $h_1$ and $h_2$ are two arbitrary functions of time. Accordingly, we have
\begin{equation}
p=\frac{\partial F}{\partial x}=h_1\,\bar{p}+2h_2\,x \quad\text{and}\quad\bar{x}=\frac{\partial F}{\partial \bar{p}}=h_1\,x \,.
\end{equation}
Noting, $ \mc{H}(\bar{x},\bar{p},t)=\widetilde{H}(\bar{x},\bar{p},t)+\pd F/\pd t $, it is then straight forward to show that
\begin{equation}
\begin{split}
\mc{H}(\bar{x},\bar{p},t)=\frac{\dot{f}h_1^2}{2m}\,\bar{p}^2+\left(\frac{4h_2^2\dot{f}}{2mh_1^2}+\frac{\dot{h_2}}{h_1^2}\right)\,\bar{x}^2+\left(\frac{4h_2\dot{f}}{2m}+\frac{\dot{h_1}}{h_1}\right)\,\bar{p}\bar{x}+\frac{h_1^2\,f}{\bar{\gamma}^2}\,V\left(\frac{\bar{x}}{\bar{\gamma}}\right)
\end{split}
\end{equation}
where $\bar{\gamma}=h_1\gamma$ and we suppressed the time-dependence of the parameters to avoid clutter. 

Choosing the arbitrary functions $h_1$ and $h_2$ as
\begin{equation}
h_1=1/\sqrt{\dot{f}}\quad\text{and}\quad h_2=m\ddot{f}/4\dot{f}^2
\end{equation}
the non-local term disappears and the time-dependence of the kinetic terms is remedied. Hence, we can write 
\begin{equation}
\mc{H}(\bar{x},\bar{p},\tau)=\frac{\bar{p}^2}{2m}+\frac{\dot{f}}{\gamma^2}\,V\left(\frac{\bar{x}\sqrt{\dot{f}}}{\gamma}\right)+\kappa \bar{x}^2
\end{equation}
with $ \kappa=m\ddot{f}^2/8\dot{f}^2+m(\dot{f}\ddot{f}-2\ddot{f}^3)/4\dot{f}$.

In conclusion, we have shown that time-rescaling in scale-invariant problems is equivalent to including an auxiliary field, which is simply given by a harmonic term. Since it also has been shown that the counterdiabatic field in transitionless quantum driving can be reduced to a  harmonic term \cite{Deffner2014}, we have that time-rescaling can indeed facilitate transitionless quantum driving.

\subsection{Scale-invariant Schr\"odinger dynamics}

The analogous problem can also be solved for Schr\"odinger dynamics. In this case, we write the quantum generating function as
\begin{equation}
\mc{F}(x,p,t)=\exp{\frac{i}{\hbar}\,\alpha(t) x^2}\exp{\frac{i}{\hbar}\,\beta(t)\{x,p\}}
\end{equation}
for two arbitrary functions $\alpha(t)$ and $\beta(t)$, and $\{x,p\}=xp+px$. Now choosing,
\begin{equation}
\beta(t)=\log(\dot{f}(t))\quad\text{and}\quad\alpha(t)=\frac{\ddot{f}(t)}{\dot{f}(t)}\,, 
\end{equation}
we obtain for $\widetilde{H}(t)=\dot{f}(t)\, H(t)$
\begin{equation}
\mc{H}(x,p,t)=\frac{p^2}{2m}+\frac{\dot{f}}{\gamma^2}\,V\left(\frac{x}{\dot{f}\gamma}\right)+\kappa_q\,x^2\,,
\end{equation}
and $\kappa_q=\dddot{f}/\dot{f}-\ddot{f}^2/\dot{f}^2-\ddot{f}^2/\dot{f}^3$. In complete analogy to the classical case, time-rescaling in the original frame is facilitated by an auxiliary harmonic term.


\reftitle{References}

\externalbibliography{yes}
\bibliography{STA_Dirac}

\begin{thebibliography}{-------}
\providecommand{\natexlab}[1]{#1}

\bibitem[Born(1927)]{Born1927}
Born, M.
\newblock Das Adiabatenprinzip in der Quantenmechanik.
\newblock {\em Z. Physik} {\bf 1927}, {\em 40},~167.
\newblock
  doi:{\changeurlcolor{black}\href{https://doi.org/10.1007/BF01400360}{\detokenize{10.1007/BF01400360}}}.

\bibitem[Messiah(1966)]{Messiah1966}
Messiah, A.
\newblock {\em Quantum Mechanics}; Vol.~II, John Wiley \& Sons: Amsterdam, The
  Netherlands,  1966.

\bibitem[Nenciu(1980{\natexlab{a}})]{Nenciu1980}
Nenciu, G.
\newblock On the adiabatic theorem of quantum mechanics.
\newblock {\em Journal of Physics A: Mathematical and General} {\bf 1980}, {\em
  13},~L15--L18.
\newblock
  doi:{\changeurlcolor{black}\href{https://doi.org/10.1088/0305-4470/13/2/002}{\detokenize{10.1088/0305-4470/13/2/002}}}.

\bibitem[Nenciu(1980{\natexlab{b}})]{Nenciu1980a}
Nenciu, G.
\newblock {On the adiabatic limit for Dirac particles in external fields}.
\newblock {\em Commun. Math. Phys.} {\bf 1980}, {\em 76},~117.
\newblock
  doi:{\changeurlcolor{black}\href{https://doi.org/10.1007/BF01212820}{\detokenize{10.1007/BF01212820}}}.

\bibitem[Nenciu(1981)]{Nenciu1981}
Nenciu, G.
\newblock Adiabatic theorem and spectral concentration.
\newblock {\em Commun. Math. Phys.} {\bf 1981}, {\em 82},~121.
\newblock
  doi:{\changeurlcolor{black}\href{https://doi.org/10.1007/BF01206948}{\detokenize{10.1007/BF01206948}}}.

\bibitem[Albash and Lidar(2018)]{Albash2018}
Albash, T.; Lidar, D.A.
\newblock Adiabatic quantum computation.
\newblock {\em Rev. Mod. Phys.} {\bf 2018}, {\em 90},~015002.
\newblock
  doi:{\changeurlcolor{black}\href{https://doi.org/10.1103/RevModPhys.90.015002}{\detokenize{10.1103/RevModPhys.90.015002}}}.

\bibitem[Young \em{et~al.}(2013)Young, Sarovar, and Blume-Kohout]{Young2013}
Young, K.C.; Sarovar, M.; Blume-Kohout, R.
\newblock Error Suppression and Error Correction in Adiabatic Quantum
  Computation: Techniques and Challenges.
\newblock {\em Phys. Rev. X} {\bf 2013}, {\em 3},~041013.
\newblock
  doi:{\changeurlcolor{black}\href{https://doi.org/10.1103/PhysRevX.3.041013}{\detokenize{10.1103/PhysRevX.3.041013}}}.

\bibitem[Gardas and Deffner(2018)]{Gardas2018}
Gardas, B.; Deffner, S.
\newblock Quantum fluctuation theorem for error diagnostics in quantum
  annealers.
\newblock {\em Sci. Rep.} {\bf 2018}, {\em 8},~17191.
\newblock
  doi:{\changeurlcolor{black}\href{https://doi.org/10.1038/s41598-018-35264-z}{\detokenize{10.1038/s41598-018-35264-z}}}.

\bibitem[Demirplak and Rice(2003)]{Demirplak2003}
Demirplak, M.; Rice, S.A.
\newblock Adiabatic Population Transfer with Control Fields.
\newblock {\em J. Chem. Phys. A} {\bf 2003}, {\em 107},~9937.
\newblock
  doi:{\changeurlcolor{black}\href{https://doi.org/10.1021/jp030708a}{\detokenize{10.1021/jp030708a}}}.

\bibitem[Demirplak and Rice(2005)]{Demirplak2005}
Demirplak, M.; Rice, S.A.
\newblock {Assisted adiabatic passage revisited}.
\newblock {\em J. Phys. Chem. B} {\bf 2005}, {\em 109},~6838.
\newblock
  doi:{\changeurlcolor{black}\href{https://doi.org/https://doi.org/10.1021/jp040647w}{\detokenize{https://doi.org/10.1021/jp040647w}}}.

\bibitem[Berry(2009)]{Berry2009}
Berry, M.
\newblock Transitionless quantum driving.
\newblock {\em J. Phys. A: Math. Theor.} {\bf 2009}, {\em 42},~365303.
\newblock
  doi:{\changeurlcolor{black}\href{https://doi.org/10.1088/1751-8113/42/36/365303}{\detokenize{10.1088/1751-8113/42/36/365303}}}.

\bibitem[del Campo(2013)]{delCampi2013}
del Campo, A.
\newblock Shortcuts to Adiabaticity by Counterdiabatic Driving.
\newblock {\em Phys. Rev. Lett.} {\bf 2013}, {\em 111},~100502.
\newblock
  doi:{\changeurlcolor{black}\href{https://doi.org/10.1103/PhysRevLett.111.100502}{\detokenize{10.1103/PhysRevLett.111.100502}}}.

\bibitem[Deffner \em{et~al.}(2014)Deffner, Jarzynski, and del
  Campo]{Deffner2014}
Deffner, S.; Jarzynski, C.; del Campo, A.
\newblock {Classical and quantum shortcuts to adiabaticity for scale-invariant
  driving}.
\newblock {\em Phys. Rev. X} {\bf 2014}, {\em 4},~021013.
\newblock
  doi:{\changeurlcolor{black}\href{https://doi.org/10.1103/PhysRevX.4.021013}{\detokenize{10.1103/PhysRevX.4.021013}}}.

\bibitem[Masuda and Nakamura(2010)]{Masuda2010}
Masuda, S.; Nakamura, K.
\newblock Fast-forward of adiabatic dynamics in quantum mechanics.
\newblock {\em Proc. R. Soc. A} {\bf 2010}, {\em 466},~1135.
\newblock
  doi:{\changeurlcolor{black}\href{https://doi.org/https://doi.org/10.1098/rspa.2009.0446}{\detokenize{https://doi.org/10.1098/rspa.2009.0446}}}.

\bibitem[Masuda and Nakamura(2011)]{Masuda2011}
Masuda, S.; Nakamura, K.
\newblock Acceleration of adiabatic quantum dynamics in electromagnetic fields.
\newblock {\em Phys. Rev. A} {\bf 2011}, {\em 84},~043434.
\newblock
  doi:{\changeurlcolor{black}\href{https://doi.org/http://dx.doi.org/10.1103/PhysRevA.84.043434}{\detokenize{http://dx.doi.org/10.1103/PhysRevA.84.043434}}}.

\bibitem[{Masuda} and {Rice}(2015)]{Masuda2014}
{Masuda}, S.; {Rice}, S.A.
\newblock {Fast-Forward Assisted STIRAP}.
\newblock {\em J. Phys. Chem. A} {\bf 2015}, {\em 119},~3479.
\newblock
  doi:{\changeurlcolor{black}\href{https://doi.org/10.1021/acs.jpca.5b00525}{\detokenize{10.1021/acs.jpca.5b00525}}}.

\bibitem[Masuda \em{et~al.}(2018)Masuda, Nakamura, and Nakahara]{Masuda2018}
Masuda, S.; Nakamura, K.; Nakahara, M.
\newblock Fast-forward scaling theory for phase imprinting on a {BEC}: creation
  of a wave packet with uniform momentum density and loading to Bloch states
  without disturbance.
\newblock {\em New J. Phys.} {\bf 2018}, {\em 20},~025008.
\newblock
  doi:{\changeurlcolor{black}\href{https://doi.org/10.1088/1367-2630/aaacea}{\detokenize{10.1088/1367-2630/aaacea}}}.

\bibitem[Chen \em{et~al.}(2010)Chen, Ruschhaupt, Schmidt, del Campo,
  Gu\'ery-Odelin, and Muga]{Chen2010}
Chen, X.; Ruschhaupt, A.; Schmidt, S.; del Campo, A.; Gu\'ery-Odelin, D.; Muga,
  J.G.
\newblock Fast Optimal Frictionless Atom Cooling in Harmonic Traps: Shortcut to
  Adiabaticity.
\newblock {\em \PRL} {\bf 2010}, {\em 104},~063002.
\newblock
  doi:{\changeurlcolor{black}\href{https://doi.org/10.1103/PhysRevLett.104.063002}{\detokenize{10.1103/PhysRevLett.104.063002}}}.

\bibitem[Torrontegui \em{et~al.}(2014)Torrontegui, Mart\'{\i}nez-Garaot, and
  Muga]{Torrontegui2014}
Torrontegui, E.; Mart\'{\i}nez-Garaot, S.; Muga, J.G.
\newblock Hamiltonian engineering via invariants and dynamical algebra.
\newblock {\em Phys. Rev. A} {\bf 2014}, {\em 89},~043408.
\newblock
  doi:{\changeurlcolor{black}\href{https://doi.org/10.1103/PhysRevA.89.043408}{\detokenize{10.1103/PhysRevA.89.043408}}}.

\bibitem[Kiely \em{et~al.}(2015)Kiely, McGuinness, Muga, and
  Ruschhaupt]{Kiely2015}
Kiely, A.; McGuinness, J.P.L.; Muga, J.G.; Ruschhaupt, A.
\newblock Fast and stable manipulation of a charged particle in a Penning trap.
\newblock {\em J. Phys. B: At. Mol. Opt. Phys.} {\bf 2015}, {\em 48},~075503.
\newblock
  doi:{\changeurlcolor{black}\href{https://doi.org/10.1088/0953-4075/48/7/075503}{\detokenize{10.1088/0953-4075/48/7/075503}}}.

\bibitem[Jarzynski \em{et~al.}(2017)Jarzynski, Deffner, Patra, and Suba\ifmmode
  \mbox{\c{s}}\else \c{s}\fi{}\ifmmode \imath \else~\i \fi{}]{Jarzynski2017}
Jarzynski, C.; Deffner, S.; Patra, A.; Suba\ifmmode \mbox{\c{s}}\else
  \c{s}\fi{}\ifmmode \imath \else~\i \fi{}, Y.b.u.
\newblock Fast forward to the classical adiabatic invariant.
\newblock {\em Phys. Rev. E} {\bf 2017}, {\em 95},~032122.
\newblock
  doi:{\changeurlcolor{black}\href{https://doi.org/10.1103/PhysRevE.95.032122}{\detokenize{10.1103/PhysRevE.95.032122}}}.

\bibitem[Torrontegui \em{et~al.}(2013)Torrontegui, Ib\'{a}\~{n}ez,
  Mart\'{\i}nez-Garaot, Modugno, del Campo, Gu\'{e}ry-Odelin, Ruschhaupt, Chen,
  and Muga]{Torrontegui2013}
Torrontegui, E.; Ib\'{a}\~{n}ez, S.; Mart\'{\i}nez-Garaot, S.; Modugno, M.; del
  Campo, A.; Gu\'{e}ry-Odelin, D.; Ruschhaupt, A.; Chen, X.; Muga, J.G.
\newblock {Shortcuts to Adiabaticity}.
\newblock {\em Adv. At. Mol. Opt. Phys.} {\bf 2013}, {\em 62},~117.
\newblock
  doi:{\changeurlcolor{black}\href{https://doi.org/10.1016/B978-0-12-408090-4.00002-5}{\detokenize{10.1016/B978-0-12-408090-4.00002-5}}}.

\bibitem[Gu\'ery-Odelin \em{et~al.}(2019)Gu\'ery-Odelin, Ruschhaupt, Kiely,
  Torrontegui, Mart\'{\i}nez-Garaot, and Muga]{Guery2019}
Gu\'ery-Odelin, D.; Ruschhaupt, A.; Kiely, A.; Torrontegui, E.;
  Mart\'{\i}nez-Garaot, S.; Muga, J.G.
\newblock Shortcuts to adiabaticity: Concepts, methods, and applications.
\newblock {\em Rev. Mod. Phys.} {\bf 2019}, {\em 91},~045001.
\newblock
  doi:{\changeurlcolor{black}\href{https://doi.org/10.1103/RevModPhys.91.045001}{\detokenize{10.1103/RevModPhys.91.045001}}}.

\bibitem[del Campo and Kim(2019)]{delCampo2019}
del Campo, A.; Kim, K.
\newblock Focus on Shortcuts to Adiabaticity.
\newblock {\em New J. Phys.} {\bf 2019}, {\em 21},~050201.
\newblock
  doi:{\changeurlcolor{black}\href{https://doi.org/10.1088/1367-2630/ab1437}{\detokenize{10.1088/1367-2630/ab1437}}}.

\bibitem[Deffner and Bonan{\c{c}}a(2020)]{Deffner2020EPL}
Deffner, S.; Bonan{\c{c}}a, M.V.S.
\newblock Thermodynamic control {\textemdash}An old paradigm with new
  applications.
\newblock {\em {EPL} (Europhysics Letters)} {\bf 2020}, {\em 131},~20001.
\newblock
  doi:{\changeurlcolor{black}\href{https://doi.org/10.1209/0295-5075/131/20001}{\detokenize{10.1209/0295-5075/131/20001}}}.

\bibitem[Alipour \em{et~al.}(2020)Alipour, Chenu, Rezakhani, and del
  Campo]{Alipour2020}
Alipour, S.; Chenu, A.; Rezakhani, A.T.; del Campo, A.
\newblock Shortcuts to {A}diabaticity in {D}riven {O}pen {Q}uantum {S}ystems:
  {B}alanced {G}ain and {L}oss and {N}on-{M}arkovian {E}volution.
\newblock {\em {Quantum}} {\bf 2020}, {\em 4},~336.
\newblock
  doi:{\changeurlcolor{black}\href{https://doi.org/10.22331/q-2020-09-28-336}{\detokenize{10.22331/q-2020-09-28-336}}}.

\bibitem[Patra and Jarzynski(2017{\natexlab{a}})]{Patra2017}
Patra, A.; Jarzynski, C.
\newblock Classical and Quantum Shortcuts to Adiabaticity in a Tilted Piston.
\newblock {\em The Journal of Physical Chemistry B} {\bf 2017}, {\em
  121},~3403--3411.
\newblock
  doi:{\changeurlcolor{black}\href{https://doi.org/10.1021/acs.jpcb.6b08769}{\detokenize{10.1021/acs.jpcb.6b08769}}}.

\bibitem[Patra and Jarzynski(2017{\natexlab{b}})]{Patra2017NJP}
Patra, A.; Jarzynski, C.
\newblock Shortcuts to adiabaticity using flow fields.
\newblock {\em New J. Phys.} {\bf 2017}, {\em 19},~125009.
\newblock
  doi:{\changeurlcolor{black}\href{https://doi.org/10.1088/1367-2630/aa924c}{\detokenize{10.1088/1367-2630/aa924c}}}.

\bibitem[Iram \em{et~al.}(2020)Iram, Dolson, Chiel, Pelesko, Krishnan,
  G{\"u}ng{\"o}r, Kuznets-Speck, Deffner, Ilker, Scott, and
  Hinczewski]{Iram2020}
Iram, S.; Dolson, E.; Chiel, J.; Pelesko, J.; Krishnan, N.; G{\"u}ng{\"o}r,
  {\"O}.; Kuznets-Speck, B.; Deffner, S.; Ilker, E.; Scott, J.G.; Hinczewski,
  M.
\newblock Controlling the speed and trajectory of evolution with
  counterdiabatic driving.
\newblock {\em Nat. Phys.} {\bf 2020}.
\newblock
  doi:{\changeurlcolor{black}\href{https://doi.org/10.1038/s41567-020-0989-3}{\detokenize{10.1038/s41567-020-0989-3}}}.

\bibitem[Deffner(2015)]{Deffner2015b}
Deffner, S.
\newblock Shortcuts to adiabaticity: suppression of pair production in driven
  Dirac dynamics.
\newblock {\em New J. Phys.} {\bf 2015}, {\em 18},~012001.
\newblock
  doi:{\changeurlcolor{black}\href{https://doi.org/10.1088/1367-2630/18/1/012001}{\detokenize{10.1088/1367-2630/18/1/012001}}}.

\bibitem[Fan \em{et~al.}(2018)Fan, Cheng, and Chen]{fan2017}
Fan, Q.Z.; Cheng, X.H.; Chen, X.
\newblock {Counter-diabatic driving for Dirac dynamics}.
\newblock  Young Scientists Forum 2017; Zhuang, S.; Chu, J.; Pan, J.W., Eds.
  International Society for Optics and Photonics, SPIE,  2018, Vol. 10710, pp.
  42 -- 47.
\newblock
  doi:{\changeurlcolor{black}\href{https://doi.org/10.1117/12.2314712}{\detokenize{10.1117/12.2314712}}}.

\bibitem[Song \em{et~al.}(2017)Song, Deng, Lamata, and Muga]{song2017}
Song, X.K.; Deng, F.G.; Lamata, L.; Muga, J.G.
\newblock Robust state preparation in quantum simulations of Dirac dynamics.
\newblock {\em Phys. Rev. A} {\bf 2017}, {\em 95},~022332.
\newblock
  doi:{\changeurlcolor{black}\href{https://doi.org/10.1103/PhysRevA.95.022332}{\detokenize{10.1103/PhysRevA.95.022332}}}.

\bibitem[Dirac(1928)]{Dirac1928}
Dirac, P.A.M.
\newblock The quantum theory of the electron.
\newblock {\em Proc. R. Soc. A} {\bf 1928}, {\em 117},~778.
\newblock
  doi:{\changeurlcolor{black}\href{https://doi.org/10.1098/rspa.1928.0023}{\detokenize{10.1098/rspa.1928.0023}}}.

\bibitem[Thaller(1956)]{Thaller1956}
Thaller, B.
\newblock {\em {The Dirac equation}}; Springer: Berlin, Germany,  1956.

\bibitem[Peskin and Schroeder(1995)]{Peskin1995}
Peskin, M.E.; Schroeder, D.V.
\newblock {\em An Introduction To Quantum Field Theory (Frontiers in Physics)};
  Westview Press: Boulder, CO, USA,  1995.

\bibitem[Pickl and D\"urr(2008)]{Pickl2008}
Pickl, P.; D\"urr, D.
\newblock On Adiabatic Pair Creation.
\newblock {\em Commun. Math. Phys.} {\bf 2008}, {\em 282},~161.
\newblock
  doi:{\changeurlcolor{black}\href{https://doi.org/10.1007/s00220-008-0530-5}{\detokenize{10.1007/s00220-008-0530-5}}}.

\bibitem[Fillion-Gourdeau \em{et~al.}(2012)Fillion-Gourdeau, Lorin, and
  Bandrauk]{Fillion2012}
Fillion-Gourdeau, F.; Lorin, E.; Bandrauk, A.D.
\newblock {L}andau-{Z}ener-{S}t\"uckelberg interferometry in pair production
  from counterpropagating lasers.
\newblock {\em Phys. Rev. A} {\bf 2012}, {\em 86},~032118.
\newblock
  doi:{\changeurlcolor{black}\href{https://doi.org/10.1103/PhysRevA.86.032118}{\detokenize{10.1103/PhysRevA.86.032118}}}.

\bibitem[Fillion-Gourdeau \em{et~al.}(2013{\natexlab{a}})Fillion-Gourdeau,
  Lorin, and Bandrauk]{Fillion2013}
Fillion-Gourdeau, F.; Lorin, E.; Bandrauk, A.D.
\newblock Resonantly Enhanced Pair Production in a Simple Diatomic Model.
\newblock {\em Phys. Rev. Lett.} {\bf 2013}, {\em 110},~013002.
\newblock
  doi:{\changeurlcolor{black}\href{https://doi.org/10.1103/PhysRevLett.110.013002}{\detokenize{10.1103/PhysRevLett.110.013002}}}.

\bibitem[Fillion-Gourdeau \em{et~al.}(2013{\natexlab{b}})Fillion-Gourdeau,
  Lorin, and Bandrauk]{Fillion2013a}
Fillion-Gourdeau, F.; Lorin, E.; Bandrauk, A.D.
\newblock {Enhanced Schwinger pair production in many-centre systems}.
\newblock {\em J. Phys. B: At. Mol. Opt. Phys.} {\bf 2013}, {\em 46},~175002.
\newblock
  doi:{\changeurlcolor{black}\href{https://doi.org/https://doi.org/10.1088/0953-4075/46/17/175002}{\detokenize{https://doi.org/10.1088/0953-4075/46/17/175002}}}.

\bibitem[Fillion-Gourdeau and MacLean(2015)]{Fillion2015}
Fillion-Gourdeau, F.; MacLean, S.
\newblock {Time-dependent pair creation and the Schwinger mechanism in
  graphene}.
\newblock {\em Phys. Rev. B} {\bf 2015}, {\em 92},~035401.
\newblock
  doi:{\changeurlcolor{black}\href{https://doi.org/10.1103/PhysRevB.92.035401}{\detokenize{10.1103/PhysRevB.92.035401}}}.

\bibitem[Villamizar and Duzzioni(2015)]{Villamizar2015}
Villamizar, D.V.; Duzzioni, E.I.
\newblock Quantum speed limit for a relativistic electron in a uniform magnetic
  field.
\newblock {\em Phys. Rev. A} {\bf 2015}, {\em 92},~042106.
\newblock
  doi:{\changeurlcolor{black}\href{https://doi.org/10.1103/PhysRevA.92.042106}{\detokenize{10.1103/PhysRevA.92.042106}}}.

\bibitem[Schmidt \em{et~al.}(2015)Schmidt, Peano, and Marquardt]{Schmidt2015}
Schmidt, M.; Peano, V.; Marquardt, F.
\newblock {Optomechanical Dirac physics}.
\newblock {\em New J. Phys.} {\bf 2015}, {\em 17},~023025.
\newblock
  doi:{\changeurlcolor{black}\href{https://doi.org/https://doi.org/10.1088/1367-2630/17/2/023025}{\detokenize{https://doi.org/10.1088/1367-2630/17/2/023025}}}.

\bibitem[Deffner and Saxena(2015)]{Deffner2015}
Deffner, S.; Saxena, A.
\newblock {Quantum work statistics of charged Dirac particles in time-dependent
  fields}.
\newblock {\em Phys. Rev. E} {\bf 2015}, {\em 92},~032137.
\newblock
  doi:{\changeurlcolor{black}\href{https://doi.org/10.1103/PhysRevE.92.032137}{\detokenize{10.1103/PhysRevE.92.032137}}}.

\bibitem[{Wehling} \em{et~al.}(2014){Wehling}, {Black-Schaffer}, and
  {Balatsky}]{Wehling2014}
{Wehling}, T.O.; {Black-Schaffer}, A.M.; {Balatsky}, A.V.
\newblock {Dirac materials}.
\newblock {\em Adv. Phys.} {\bf 2014}, {\em 63},~1.
\newblock
  doi:{\changeurlcolor{black}\href{https://doi.org/10.1080/00018732.2014.927109}{\detokenize{10.1080/00018732.2014.927109}}}.

\bibitem[Faisal(2011)]{Faisal2011}
Faisal, F.H.M.
\newblock {Adiabatic solutions of a Dirac equation of a new class of
  quasi-particles and high harmonic generation from them in an intense
  electromagnetic field}.
\newblock {\em J. Phys. B: At. Mol. Opt. Phys.} {\bf 2011}, {\em 44},~111001.
\newblock
  doi:{\changeurlcolor{black}\href{https://doi.org/10.1088/0953-4075/44/11/111001}{\detokenize{10.1088/0953-4075/44/11/111001}}}.

\bibitem[Bernardo(2020)]{Bernardo2020}
Bernardo, B.d.L.
\newblock Time-rescaled quantum dynamics as a shortcut to adiabaticity.
\newblock {\em Phys. Rev. Research} {\bf 2020}, {\em 2},~013133.
\newblock
  doi:{\changeurlcolor{black}\href{https://doi.org/10.1103/PhysRevResearch.2.013133}{\detokenize{10.1103/PhysRevResearch.2.013133}}}.

\bibitem[de~Castro(2003)]{deCastro2003}
de~Castro, A.S.
\newblock Bound states by a pseudoscalar Coulomb potential in one-plus-one
  dimensions.
\newblock {\em Phys. Lett. A} {\bf 2003}, {\em 318},~40.
\newblock
  doi:{\changeurlcolor{black}\href{https://doi.org/http://dx.doi.org/10.1016/j.physleta.2003.09.029}{\detokenize{http://dx.doi.org/10.1016/j.physleta.2003.09.029}}}.

\bibitem[Solomon(2010)]{Solomon2010}
Solomon, D.
\newblock {An exact solution of the Dirac equation for a time-dependent
  Hamiltonian in 1-1 dimension space-time}.
\newblock {\em Can. J. Phys.} {\bf 2010}, {\em 88},~137.
\newblock
  doi:{\changeurlcolor{black}\href{https://doi.org/10.1139/P10-006}{\detokenize{10.1139/P10-006}}}.

\bibitem[Haouat and Chetouani(2007)]{Haouat2007}
Haouat, S.; Chetouani, L.
\newblock {The (1+1)-Dimensional Dirac Equation With Pseudoscalar Potentials:
  Path Integral Treatment}.
\newblock {\em Int. J. Theo. Phys.} {\bf 2007}, {\em 46},~1528.
\newblock
  doi:{\changeurlcolor{black}\href{https://doi.org/10.1007/s10773-006-9290-1}{\detokenize{10.1007/s10773-006-9290-1}}}.

\bibitem[Haouat and Chetouani(2008)]{Haouat2008}
Haouat, S.; Chetouani, L.
\newblock {The (1+1)-dimensional Dirac equation with pseudoscalar potentials:
  quasi-classical approximation}.
\newblock {\em Phys. Scri.} {\bf 2008}, {\em 78},~065005.
\newblock
  doi:{\changeurlcolor{black}\href{https://doi.org/https://doi.org/10.1088/0031-8949/78/06/065005}{\detokenize{https://doi.org/10.1088/0031-8949/78/06/065005}}}.

\bibitem[Gardas \em{et~al.}(2016)Gardas, Deffner, and Saxena]{Gardas2016PRA}
Gardas, B.; Deffner, S.; Saxena, A.
\newblock Repeatability of measurements: Non-Hermitian observables and quantum
  Coriolis force.
\newblock {\em Phys. Rev. A} {\bf 2016}, {\em 94},~022121.
\newblock
  doi:{\changeurlcolor{black}\href{https://doi.org/10.1103/PhysRevA.94.022121}{\detokenize{10.1103/PhysRevA.94.022121}}}.

\bibitem[Leibfried \em{et~al.}(2003)Leibfried, Blatt, Monroe, and
  Wineland]{Leibfried2003}
Leibfried, D.; Blatt, R.; Monroe, C.; Wineland, D.
\newblock Quantum dynamics of single trapped ions.
\newblock {\em Rev. Mod. Phys.} {\bf 2003}, {\em 75},~281--324.
\newblock
  doi:{\changeurlcolor{black}\href{https://doi.org/10.1103/RevModPhys.75.281}{\detokenize{10.1103/RevModPhys.75.281}}}.

\bibitem[Lamata \em{et~al.}(2007)Lamata, Le\'on, Sch\"atz, and
  Solano]{Lamata2007}
Lamata, L.; Le\'on, J.; Sch\"atz, T.; Solano, E.
\newblock Dirac Equation and Quantum Relativistic Effects in a Single Trapped
  Ion.
\newblock {\em Phys. Rev. Lett.} {\bf 2007}, {\em 98},~253005.
\newblock
  doi:{\changeurlcolor{black}\href{https://doi.org/10.1103/PhysRevLett.98.253005}{\detokenize{10.1103/PhysRevLett.98.253005}}}.

\bibitem[Gerritsma \em{et~al.}(2010)Gerritsma, Kirchmair, Z{\"a}hringer,
  Solano, Blatt, and Roos]{Gerritsma2010}
Gerritsma, R.; Kirchmair, G.; Z{\"a}hringer, F.; Solano, E.; Blatt, R.; Roos,
  C.F.
\newblock Quantum simulation of the Dirac equation.
\newblock {\em Nature} {\bf 2010}, {\em 463},~68--71.
\newblock
  doi:{\changeurlcolor{black}\href{https://doi.org/10.1038/nature08688}{\detokenize{10.1038/nature08688}}}.

\bibitem[Muga \em{et~al.}(2016)Muga, Sim{\'{o}}n, and Tobalina]{Muga2016NJP}
Muga, J.G.; Sim{\'{o}}n, M.A.; Tobalina, A.
\newblock How to drive a Dirac system fast and safe.
\newblock {\em New J. Phys.} {\bf 2016}, {\em 18},~021005.
\newblock
  doi:{\changeurlcolor{black}\href{https://doi.org/10.1088/1367-2630/18/2/021005}{\detokenize{10.1088/1367-2630/18/2/021005}}}.

\bibitem[Floquet(1883)]{Floquet1883}
Floquet, G.
\newblock Sur les \'equations diff\'erentielles lin\'eaires \`a coefficients
  p\'eriodiques.
\newblock {\em Annales scientifiques de l'\'Ecole Normale Sup\'erieure} {\bf
  1883}, {\em 2e s{\'e}rie, 12},~47--88.
\newblock
  doi:{\changeurlcolor{black}\href{https://doi.org/10.24033/asens.220}{\detokenize{10.24033/asens.220}}}.

\bibitem[Lindner \em{et~al.}(2011)Lindner, Refael, and Galitski]{Lindner2011}
Lindner, N.H.; Refael, G.; Galitski, V.
\newblock Floquet topological insulator in semiconductor quantum wells.
\newblock {\em Nature Physics} {\bf 2011}, {\em 7},~490--495.
\newblock
  doi:{\changeurlcolor{black}\href{https://doi.org/10.1038/nphys1926}{\detokenize{10.1038/nphys1926}}}.

\bibitem[Shirley(1965)]{Shirley1965}
Shirley, J.H.
\newblock Solution of the Schr\"odinger Equation with a Hamiltonian Periodic in
  Time.
\newblock {\em Phys. Rev.} {\bf 1965}, {\em 138},~B979--B987.
\newblock
  doi:{\changeurlcolor{black}\href{https://doi.org/10.1103/PhysRev.138.B979}{\detokenize{10.1103/PhysRev.138.B979}}}.

\bibitem[Dittrich \em{et~al.}(1998)Dittrich, H\"anggi, Ingold, Kramer, Schön,
  and Zwerger]{Hanggi1998}
Dittrich, T.; H\"anggi, P.; Ingold, G.L.; Kramer, B.; Schön, G.; Zwerger, W.
\newblock {\em Quantum Transport and Dissipation}; Wiley-VCH,  1998.

\bibitem[Ho and Gong(2012)]{Ho2012}
Ho, D.Y.H.; Gong, J.
\newblock Quantized Adiabatic Transport In Momentum Space.
\newblock {\em Phys. Rev. Lett.} {\bf 2012}, {\em 109},~010601.
\newblock
  doi:{\changeurlcolor{black}\href{https://doi.org/10.1103/PhysRevLett.109.010601}{\detokenize{10.1103/PhysRevLett.109.010601}}}.

\bibitem[Bomantara \em{et~al.}(2016)Bomantara, Raghava, Zhou, and Gong]{bm16a}
Bomantara, R.W.; Raghava, G.N.; Zhou, L.; Gong, J.
\newblock Floquet topological semimetal phases of an extended kicked Harper
  model.
\newblock {\em Phys. Rev. E} {\bf 2016}, {\em 93},~022209.
\newblock
  doi:{\changeurlcolor{black}\href{https://doi.org/10.1103/PhysRevE.93.022209}{\detokenize{10.1103/PhysRevE.93.022209}}}.

\bibitem[Bomantara and Gong(2016)]{bm16b}
Bomantara, R.W.; Gong, J.
\newblock Generating controllable type-II Weyl points via periodic driving.
\newblock {\em Phys. Rev. B} {\bf 2016}, {\em 94},~235447.
\newblock
  doi:{\changeurlcolor{black}\href{https://doi.org/10.1103/PhysRevB.94.235447}{\detokenize{10.1103/PhysRevB.94.235447}}}.

\bibitem[{Carinena} \em{et~al.}(1988){Carinena}, {Ibort}, and
  {Lacomba}]{Carinena1988}
{Carinena}, J.F.; {Ibort}, L.A.; {Lacomba}, E.A.
\newblock {Time Scaling as an Infinitesimal Canonical Transformation}.
\newblock {\em Celestial Mechanics} {\bf 1988}, {\em 42},~201--213.
\newblock
  doi:{\changeurlcolor{black}\href{https://doi.org/10.1007/BF01232957}{\detokenize{10.1007/BF01232957}}}.

\bibitem[Goldstein(1980)]{Goldstein1980}
Goldstein, H.
\newblock {\em Classical Mechanics}; Addison-Wesley,  1980.

\end{thebibliography}

\end{document}